\documentclass[pra,twocolumn,showpacs,groupaddress]{revtex4-1} 
\usepackage{graphicx,amsfonts,amsmath,amssymb} 
\usepackage{pifont}
\usepackage[toc,page]{appendix} 
\usepackage{epsfig}  
\usepackage{epstopdf}
\usepackage{times}
\usepackage{txfonts}
\newcommand{\tickYes}{\hspace{1pt}\ding{51}}
\newcommand{\tickNo}{\hspace{1pt}\ding{55}} 

\begin{document}
\bibliographystyle{h-physrev3}
\title{Stability and structure of an anisotropically trapped dipolar Bose-Einstein condensate:  angular and linear rotons}
\author{A. D. Martin}
\author{P. B. Blakie}
\affiliation{Jack Dodd Centre for Quantum Technology, Department of Physics, University of Otago, Dunedin, 9016, New Zealand}

\date{\today}
\begin{abstract}
We study theoretically Bose-Einstein condensates with polarized dipolar interactions in anisotropic traps. We map the parameter space by varying the trap frequencies and dipolar interaction strengths and find an irregular-shaped region of parameter space in which density-oscillating condensate states occur, with maximum density away from the trap center. These density-oscillating states may be biconcave (red-blood-cell-shaped), or have two or four peaks. For all trap frequencies, the condensate becomes unstable to collapse for sufficiently large dipole interaction strength. The collapse coincides with the softening of  an elementary excitation. When the condensate mode is  density-oscillating, the character of the softening excitation is related to the structure of the condensate. We classify these excitations by linear and angular characteristics. We also find excited solutions to the Gross-Pitaevskii equation, which are always unstable.

\end{abstract}

 \maketitle 

\section{Introduction}
Dipolar Bose-Einstein condensates (BECs) have been produced using a variety of atoms with magnetic dipoles \cite{Griesmaier_PRL_2005, Beaufils_PRA_2008,Lu_PRL_2011,Aikawa_PRL_2012}. In this system the constituent atoms interact via a long-ranged and anisotropic dipole-dipole interaction (DDI), which has been identified as a source of interesting new effects in the degenerate regime,  such as supersolidity \cite{CSansone_PRL_2010,Pollet_PRL_2010,Chan_PRA_2010,He_PRA_2011},  solitons \cite{Nath_PRL_2009} and rotonic excitations \cite{Santos_PRL_2003,Fischer_PRA_2006}. 

To realize these novel effects, the DDI must be significant compared to the (short-range) $s$-wave interaction between atoms. 
For  $^{52}$Cr  (with dipole moment $\mu_m\approx 6 \mu_B$) the dipolar interaction has been made dominant by using a Feshbach resonance to suppress the $s$-wave interaction  \cite{Lahaye_Nature_2007, Koch_NaturePhys_2008}. With the recent realization of dipolar BECs of $^{164}$Dy \cite{Lu_PRL_2011} ($\mu_m\approx 10 \mu_B$), and $^{168}$Er \cite{Aikawa_PRL_2012} ($\mu_m\approx 7 \mu_B$)
there is now a rich and diverse set of systems for experimental investigation. 
Furthermore, progress towards the production of degenerate  polar molecules \cite{Ni_Science_2008,Aikawa_NewJPhys_2009,Lercher_EurPhysJD_2011,McCarron_PRA_2011} with large electric dipole moments promises an exciting future in this research  field.

In experiments the  dipoles are typically polarized by an external field, which induces  anisotropic (magnetostrictive) deformations  in the trap \cite{Goral_PRA_2000,Eberlein_PRA_2005,Bailie_PRA_2012} and during ballistic expansion  \cite{Giovanazzi_JOptB_2003}. Because the DDI has an attractive component (dipoles in a head to tail configuration attract each other), an important consideration is in what circumstances the condensate is mechanically stable from collapsing to a high density state  \cite{Lahaye_PRL_2008}. Experimental \cite{Koch_NaturePhys_2008,Muller_PRA_2011}  and theoretical \cite{Ronen_PRL_2007,Wilson_PRA_2009,Lu_PRA_2010, Bisset_PRA_2011} studies have shown that this stability is highly dependent upon the trap geometry.  
 The approach to instability of the condensate is associated with the softening of elementary excitations \cite{Goral_PRA_2002, Ronen_PRA_2006}. 
When the condensate is tightly confined in the direction that the dipoles are polarized, uniform studies predict that high momentum modes will soften \cite{Santos_PRL_2003,Rosenkranz_unpublished_2012}, causing a roton-like dip in the dispersion relation (similar to that observed in superfluid He \cite{Landau_JPhysUSSR_1947,Feynman_PR_1954}). In trapped dipolar BECs, the spectrum of elementary excitations is discrete and the excitation modes are not  momentum eigenstates, however calculations have verified that modes   with high momentum components soften \cite{Wilson_PRL_2010, Wilson_PRA_2011, Zaman_PRA_2011,Ronen_PRL_2007, Wilson_PRA_2009, Wilson_PRL_2008, Ticknor_PRL_2011,Blakie_PRA_2012}.

Another important feature of dipolar BECs is the emergence of structured or \textit{density-oscillating} ground states in which the peak density does not occur at the trap center. These states have been extensively studied in the cylindrically symmetric case where the condensate can develop a biconcave (red-blood-cell) \cite{Ronen_PRL_2007,Wilson_PRA_2011,Bisset_PRA_2012} or dumbbell \cite{Lu_PRA_2010} shaped density profile. Interestingly, it was shown in \cite{Ronen_PRL_2007} that when a biconcave condensate approaches mechanical instability an \textit{angular roton} (mode with non-zero angular momentum) softens, whereas for the normal (non-density-oscillating) condensate the mode that softens is purely radial in character (a \textit{radial roton}).

A study by Dutta \textit{et al.} \cite{Dutta_PRA_2007} considered the structure of the condensate in the more general case of a fully anisotropic trap, and found an array of different density-oscillating ground states. That study motivates the work we report here, where our aim is to investigate the nature of the excitations in the fully anisotropic trap, particularly those that soften in the approach to mechanical instability. Indeed, our work extends that of \cite{Ronen_PRL_2007} to show that a variety of rotonic modes, which can be broadly classified by angular or linear characteristics,  emerge in dipolar BECs. These rotons usually reveal some character of the underlying condensate, particularly if it is in a density-oscillating state. We note that the nature of the softening modes has been suggested as a probe of collapse in \cite{Wilson_PRA_2009}. During our study we also found that our ground state phase diagram was inconsistent with the results   in Ref.~\cite{Dutta_PRA_2007}.

The remainder of the paper is organised as follows: in Section \ref{Sec_Formalism}, we outline our model and methods of solution;  in Sections \ref{Sec_GSresults}-\ref{sec:gstypes} we discuss the structure and stability of dipolar BEC ground states; in Section \ref{sec:dutta} we compare our results to those in Ref.~\cite{Dutta_PRA_2007};  and in Section \ref{Sec_BdGresults} we discuss the nature of the elementary excitations whose softening accompanies the collapse. We conclude in Section \ref{Sec_Conclusions}.

\section{Formalism}\label{Sec_Formalism}
We consider a BEC of $N$ bosons trapped in an external potential
\begin{equation}
V_{\mbox{\scriptsize ext}}(\mathbf{r})=\frac{1}{2}{m}\left(\omega_x^2 x^2 +\omega_y^2 y^2+\omega_z^2 z^2 \right),
\end{equation} where the trap frequencies $\omega_{x,y,z}$ may be distinct.
We take the atoms to have magnetic dipole moment of magnitude $\mu_m$ polarized by an external field along the $z$ direction. The resulting DDI potential between atoms is of the form \cite{Lahaye_RepProgPhys_2009}
\begin{equation}
V_{\mathrm{dd}}(\mathbf{r})=\frac{\mu_0\mu_m^2}{4\pi}\frac{1-3\cos^2\theta}{r^3}\label{Eq_DDpol},
\end{equation} 
where $\mu_0$ is the permeability of free space and $\theta$ is the angle between $\mathbf{r}$ and and the $z$ axis. In general the effective low energy interaction for dipolar atoms also includes a contact interaction, i.e.,
\begin{equation}
V_{\mathrm{int}}(\mathbf{r})=\frac{4\pi a_s\hbar^2}{m}\delta(\mathbf{r})+V_{\mathrm{dd}}(\mathbf{r}),
\end{equation}
where   $a_s$ is the $s$-wave scattering length. However, here we focus on the purely dipolar case with $a_s=0$, as can be arranged in experiments by use of Feshbach resonances (e.g.~see \cite{Koch_NaturePhys_2008}).

\subsection{Condensate mode}\label{Sec_condensate}
The equilibrium unit-normalized mode $\psi(\mathbf{r})$ for the condensate is determined by the non-local  Gross-Pitaevskii equation (GPE) \cite{Goral_PRA_2000}:
\begin{align}
\mu\psi(\mathbf{r})& =[H_0 +N\int d\mathbf{r^{\prime}} \,V_{\mbox{\scriptsize int}}(\mathbf{r-r^{\prime}})|\psi(\mathbf{r^{\prime}})|^2]\psi(\mathbf{r}) ,\label{Eq_GPE}\\
&\equiv\mathcal{L}_{\mathrm{GP}}\psi(\mathbf{r}),
\end{align}
where $\mu$ is the condensate chemical potential,
\begin{align}
H_0&=-\frac{\hbar^2}{2m}\nabla^2+ V_{\mbox{\scriptsize ext}}(\mathbf{r}),  
\end{align}
is the single particle Hamiltonian,  and the last (convolution) term in Eq.~(\ref{Eq_GPE}) describes the direct (Hartree) interaction of condensate atoms with themselves.
We solve for self-consistent solutions of Eq.\ (\ref{Eq_GPE}) using a Newton Krylov algorithm, described in Appendix \ref{App_NK}.

\subsection{Elementary excitations: Bogoliubov-de Gennes equations}
Linearizing about the GPE solution  $\psi(\mathbf{r})$ the time-dependence of the condensate \footnote{The time-dependent Gross-Pitaevskii equation is obtained by making the replacement $\mu\to i\hbar\frac{\partial}{\partial t}$ in Eq.\ (\ref{Eq_GPE}).} is given by 
\begin{equation}
\Psi(\mathbf{r},t)=\Big\{\psi(\mathbf{r})+\frac{1}{\sqrt{N}}\sum_j\left[c_ju_j(\mathbf{r})e^{-i\omega_j t}+c_j^*v_j^*(\mathbf{r})e^{i\omega_j t}\right]\Big\}e^{-i\mu t},\label{linexpn}
\end{equation}
where the $c_j$'s are constants (e.g.~see \cite{Morgan_PRA_1998}).
The  modes $\{u_j,v_j\}$, with respective frequencies $\omega_j$, are the elementary (quasi-particle) excitations of the condensate, described by the Bogoliubov-de Gennes (BdG) equations 
\begin{equation}
\mathcal{L}\left(\begin{array}{c}u_i\\ v_i\end{array}\right)=\hbar\omega_i\left(\begin{array}{c}u_i\\ v_i\end{array}\right),\label{Eq_BdG}
\end{equation}
where
\begin{equation} \mathcal{L}=\left(\begin{array}{cc}
\mathcal{L}_{\mathrm{GP}}-\mu+Q\chi_1Q& Q\chi_2Q^*\\
Q^*\chi_2^*Q &-(\mathcal{L}_{\mathrm{GP}}-\mu-Q^*\chi_1Q^*)
 \end{array} \right).\label{Eq_LBdG}\end{equation}
The  operators  $\chi_1$ and $\chi_2$  
\begin{eqnarray} 
\chi_1f(\mathbf{r})=&N\int d\mathbf{r^\prime}\psi^*(\mathbf{r^\prime})V_{\mbox{\scriptsize int}}(\mathbf{r-r^\prime}) f(\mathbf{r^\prime})\psi(\mathbf{r}),\\
\chi_2f(\mathbf{r})=&N\int d\mathbf{r^\prime}\psi(\mathbf{r^\prime})V_{\mbox{\scriptsize int}}(\mathbf{r-r^\prime}) f(\mathbf{r^\prime})\psi(\mathbf{r}),
\end{eqnarray}
describe the exchange interactions between the condensate and excited modes, 
and 
$Q$ is the projection onto the subspace orthogonal to the condensate:
\begin{equation}
Qf(\mathbf{r})=f(\mathbf{r})-\psi(\mathbf{r})\int d \mathbf{r^{\prime}} \psi^*(\mathbf{r^{\prime}})f(\mathbf{r^{\prime}}).
\end{equation}
The method of solution of the BdG equations (\ref{Eq_BdG}) is explained in Appendix \ref{App_BdG} .
\section{Results}\label{resultssec}
\begin{figure}[tbph!]
\centering
\includegraphics[width=\columnwidth]{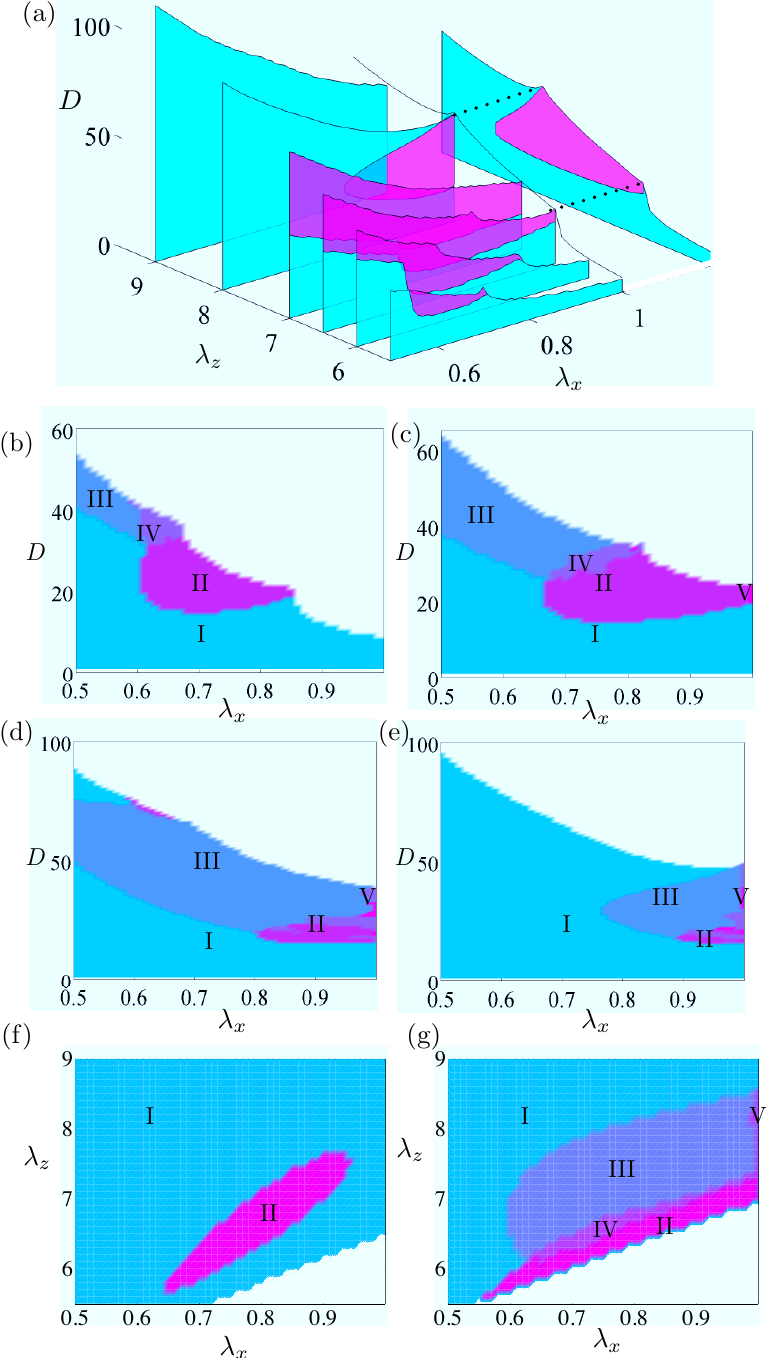}
\caption{(Color online) (a) Stability plot of a dipolar condensate in an anisotropic trap as a function of the anisotropy parameters $\lambda_x$ and $\lambda_z$ and dipole strength $D$. In the dark-shaded (magenta) region there exist stable density-oscillating condensates (with the maximum density away from the center of the condensate). The light-shaded (cyan) regions are normal ground states with maximum density at trap center. The slice in the $\lambda_z$ direction, corresponding to cylindrical symmetry $\lambda_x=1$, is displaced from the axes to improve visibility and reproduces Fig.\ 1 of Ref.\ \cite{Ronen_PRL_2007}. (b)-(e) Stability plots for constant $\lambda_z$=6 (b), 6.5 (c), 7 (d) and 8 (e) with the type of state labelled I-V according to the classification given in Sec.~\ref{sec:gstypes}. (f)-(g) Stability plots for constant $D=15$ (f) and 30 (g).
} \label{Fig_Structure3}
\end{figure}
Here we consider a harmonic trap which is, in general, fully anisotropic, i.e., the three trap frequencies are different. In our results we use $\omega_y$ and $l_y=\sqrt{\hbar/m\omega_y}$ as the reference frequency and length scale, and characterize the trap by the two anisotropy parameters 
\begin{align}
\lambda_z&=\omega_z/\omega_y,\\
\lambda_x&=\omega_x/\omega_y.
\end{align}
It is important to note that even if $\lambda_x=\lambda_z=1$ (spherical trap), the system is only cylindrically symmetric about the $z$ axis because the dipoles are polarized along this direction.  Thus, whenever  $\lambda_x\ne1$ the system breaks cylindrical symmetry, and requires the fully three-dimensional numerical solution.
We also introduce the dimensionless interaction parameter
\begin{equation}
D= \frac{\mu_0\mu_m^2}{4\pi}\frac{Nm}{\hbar^2l_y},
\end{equation}
which is the same as that used  in \cite{Ronen_PRA_2006,Ronen_PRL_2007} in the cylindrically symmetric limit. Any ground state is uniquely specified by the parameter tuple $(\lambda_x,\lambda_z,D)$.

It is worth noting that the results presented in \cite{Dutta_PRA_2007} made the unconventional choice of dipoles polarized along the $y$-direction and used the $x$-oscillator parameters to define their units. Our results can be compared to theirs by cyclically permuting the coordinates, i.e. $\lambda_x^{\mathrm{our}}\leftrightarrow\lambda_z^{\mathrm{their}}$,  $\lambda_z^{\mathrm{our}}\leftrightarrow\lambda_y^{\mathrm{their}}$.

\subsection{Ground state stability}\label{Sec_GSresults}

Our key results summarizing the stability diagram are present in Fig.~\ref{Fig_Structure3}(a). The shaded regions in this plot indicate where  a stable ground state can be located, with dark shading used to indicate the region in which a density-oscillating state occurs (i.e., where the condensate does not have its peak density at trap center).  Our results for the case of cylindrically symmetric confinement $\lambda_x=1$ correspond to those in Ref.~\cite{Ronen_PRL_2007}, including that the density-oscillating states are biconcave. Dipolar stability is highly dependent upon geometry, as is revealed in Fig.~\ref{Fig_Structure3}(a). A general trend of Fig.~\ref{Fig_Structure3}(a) is that stability increases as $\lambda_z$ increases and $\lambda_x$ decreases. This can be understood because the DDI is attractive for dipoles in a head-to-tail configuration ($z$ separation) and repulsive for dipoles in a side-by-side configuration ($xy$-plane separation). Thus increasing $\lambda_z$ (tightening $z$ confinement) reduces the number of dipoles in the destabilizing attractive configuration, while decreasing $\lambda_x$ (loosening $x$ confinement) increases the number of dipoles in the stabilizing repulsive configuration. 

\subsection{Ground state types}\label{sec:gstypes}
\begin{figure}[tbph!]
\centering
\includegraphics[width=\columnwidth]{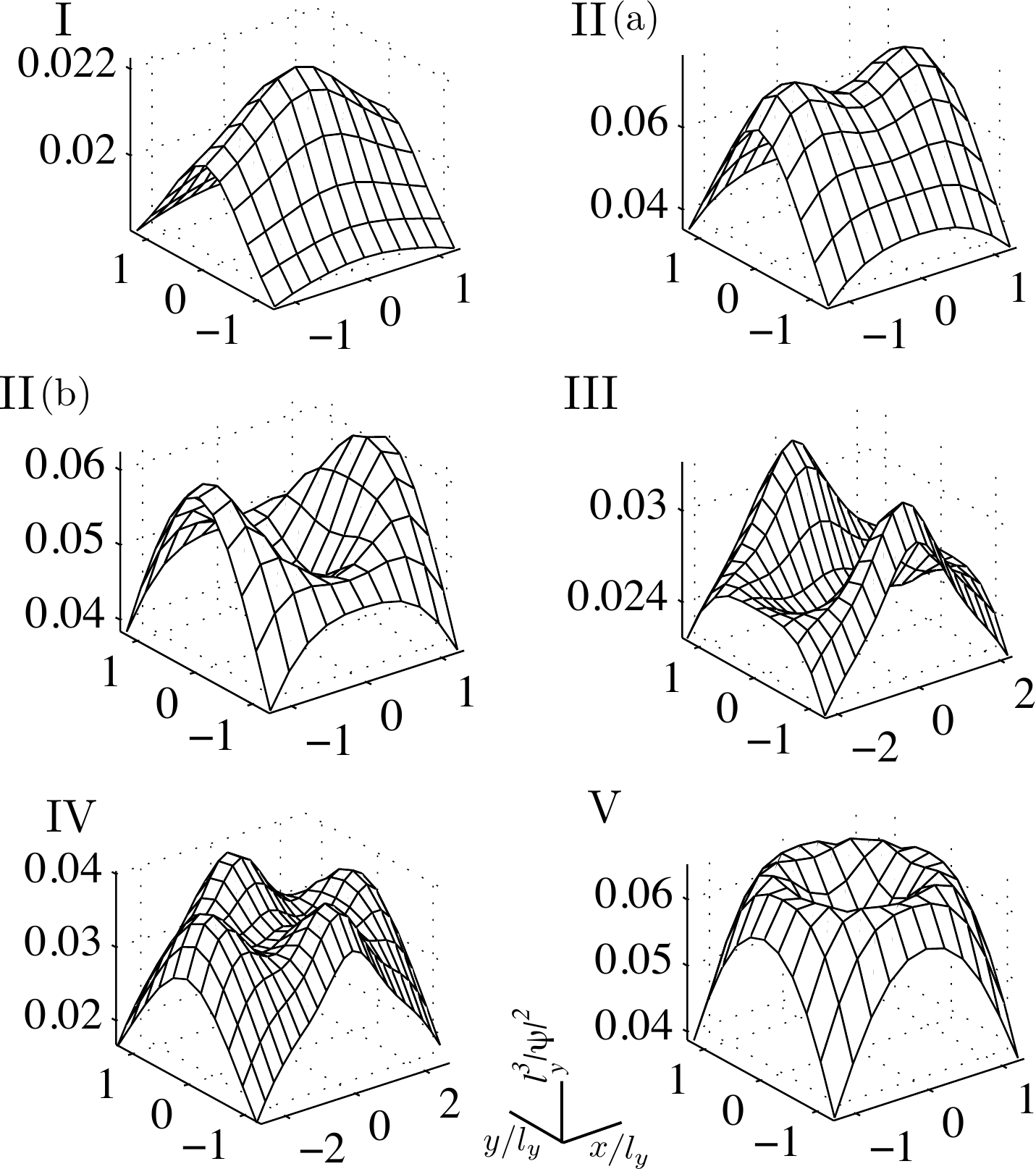}
\caption{Different types of ground state density profiles. Density slice of the condensate  in the $xy$-plane, focusing on central region where density oscillations can occur. Solutions give examples of the various condensate types (I-V) and are obtained for the parameters I $(\lambda_x,\lambda_z,D)=(0.65,9,89.4)$; II(a) $(0.85,6,20.3)$; II(b) $(0.9,6.5,26.3)$; III $(0.6,6.5,50.1)$; IV $(0.65,6,37.8)$; V $(1,6.5,24)$. See text in Sec.~\ref{sec:gstypes} for a discussion of the ground state types. 
} 
\label{Fig_StructureZoom}
\end{figure}
 
We find that a density-oscillating condensate can exhibit a range of different shapes, however in the region of interest the $z$ confinement is sufficiently tight that the non-trivial density features occur in the $xy$-plane. Following \cite{Dutta_PRA_2007} we categorize the ground state solutions by the labels I-V as follows, with reference to examples in Fig.~\ref{Fig_StructureZoom}:
\begin{enumerate}
\item[type-I] \textit{Normal}  (non-density-oscillating) condensate with peak density at the origin. 
\item[type-II] Density-oscillating condensate with two-peaks in the $x$-direction. This can either be a simple two-peaked structure [e.g.~Fig.~\ref{Fig_StructureZoom} II(a)], or two  peaks  with a biconcave crater [e.g.~Fig.~\ref{Fig_StructureZoom} II(b)].
\item[type-III] Density-oscillating condensate with two-peaks in the $y$-direction. Like type-II, this case could be a simple two-peaked case (not shown) or with a biconcave crater.
\item[type-IV] Density-oscillating condensate with four peaks (two peaks along both the $x$ and $y$-directions).
\item[type-V] Density-oscillating condensate that is biconcave (with no additional peaks).
\end{enumerate}
 
\begin{figure}[tbph!]
\centering
\includegraphics[width=\columnwidth]{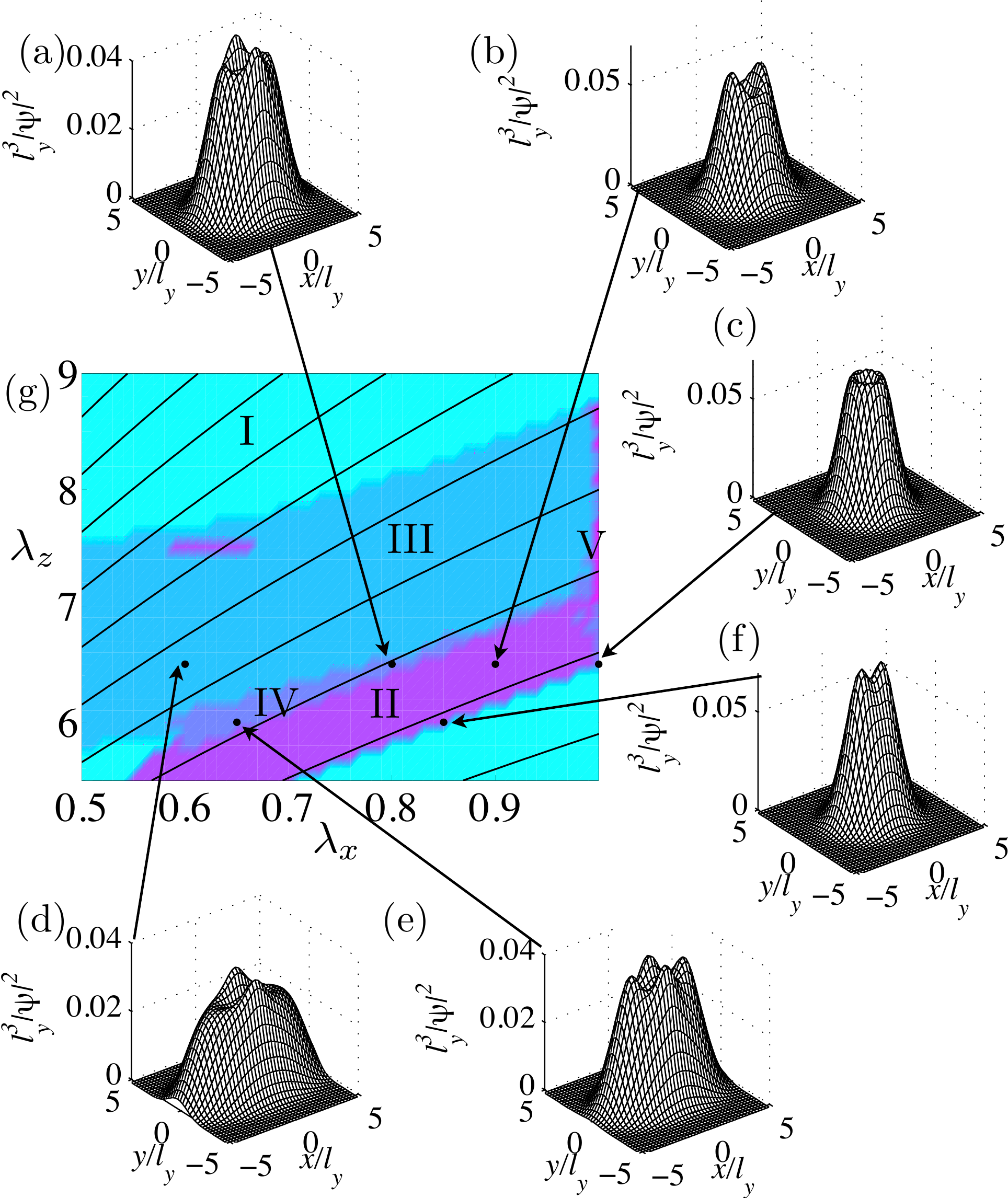}
\caption{(Color online) (a)-(f) Condensate density slices in the $xy$-plane for the parameters: (a) $(\lambda_x,\lambda_z,D)=(0.8,6.5,35.1)$; (b) $(0.9,6.5,26.3)$; (c) $(1,6.5,24)$;  (d) $(0.6,6.5,50.1)$; (e) $(0.65,6,37.8)$; (f) $(0.85,6,20.3)$;
(g) Projection of upper surface (large $D$) ground state types in Fig.\ \ref{Fig_Structure3}(a) onto the $\lambda_x\lambda_z$-plane. The density-oscillating (dark-shaded) region is divided according to the type of ground state therein. The light-shaded (cyan) regions have no stable density-oscillating ground states for any value of $D$. For reference, contours of constant $\Lambda=\omega_z/\sqrt{\omega_x\omega_y}$ are shown (thick lines).  } \label{Fig_Structure2}
\end{figure}

In Fig.~\ref{Fig_Structure2}(g) we present a projection (bird's-eye view) of the stability diagram [Fig.~\ref{Fig_Structure3}(a)] onto the $\lambda_x\lambda_z$-plane, indicating the different types of ground states which occur.
 Normal (type-I)  condensates are found in the light shaded region for any value of $D$.
Dark shaded regions indicate if a stable density-oscillating state exists for any value of $D$ in that trap geometry. The darkly shaded region is sub-divided according to the type of density-oscillating state (types II-V) at the largest dipole strength $D$ for which a stable ground state exists at that value of $\lambda_x$, $\lambda_z$. 

Contours of a generalized anisotropy $\Lambda=\omega_z/\sqrt{\omega_x\omega_y}$ are shown in Fig.~\ref{Fig_Structure2}(g). This parameter is a measure of the confinement along the direction that dipoles are polarized relative to the geometric mean confinement in the $xy$-plane (motivated by the discussion of trap effects on stability at the end of Sec.~\ref{Sec_GSresults}). We note that $\Lambda$  tends to qualitatively characterize how some boundaries of the density-oscillating region develop as the trap geometry changes. 

Figures  \ref{Fig_Structure3} and \ref{Fig_Structure2}(g)  reveal the dominant role that the type-II and III states have in the density-oscillating region. Generally  type-II states are favored at lower values of $\lambda_z$ and $D$ and  type-III states    at higher values of $\lambda_z$ and $D$. Intricate structures (lobes) in the shape of the density-oscillating condensate region arises in parameter regimes where both condensate types are present [e.g.~Figs.~\ref{Fig_Structure3}(b) and (c)]. Type-IV and V condensates are less common, with type-IV states emerging in the transition between the type-II and III regions, and type-V states occurring near cylindrical symmetry (i.e.~$\lambda_x\approx1$). As the system becomes more anisotropic [$\lambda_x<0.6$, see Fig.~\ref{Fig_Structure2}(g)] the type-II and III  sub-regions become distinct in $\lambda_x\lambda_z$-space (i.e.~separated by a normal type-I region) and appear to emerge as two separate branches. 
We note that our results for $\lambda_x<1$ can be mapped onto solutions for $\lambda_x>1$ by exchanging $x$ and $y$ coordinates (e.g.~so that type-II states become type-III etc.) and scaling the interaction parameter. This reveals that the type-II and III branches cross at $\lambda_x=1$.

\subsection{Comparison to Dutta \emph{ et al.}\ \cite{Dutta_PRA_2007}} \label{sec:dutta}
\begin{figure}[tbp]
\centering
\includegraphics[width=\columnwidth]{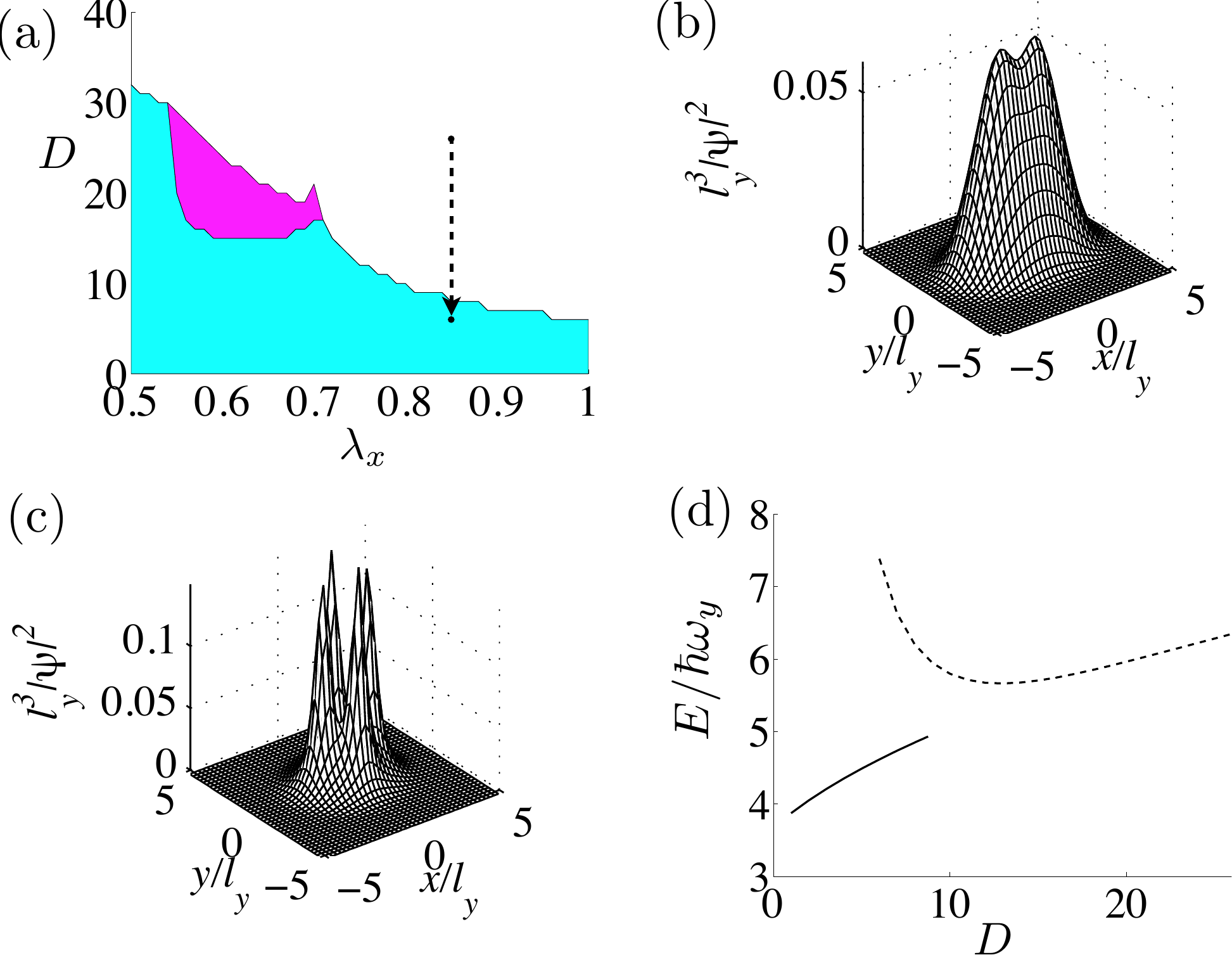}
\caption{(a) Stability plot of a dipolar condensate  as a function of $\lambda_x$, with $\lambda_z=5.5$. Two-peaked ground states are indicated by the dark (magenta) region. The dotted line indicates the plotting range of the energy of unstable four-peaked solutions shown in (d). (b) Stable two-peaked solution for $(\lambda_x,\lambda_z,D)=(0.65,5.5,20)$. (c) Unstable four-peaked GPE excitation for  $(0.85,5.5,20)$. (d) Energy per particle of type-I ground state (solid line) and unstable, four-peaked (type-IV), GPE excitation (dotted line) as a function of  $D$ for the same trap parameters as in (c).} \label{Fig_Comparison}
\end{figure}
The parameter regime in Figs.~\ref{Fig_Structure3} and \ref{Fig_Structure2} is similar to that studied by Dutta \emph{ et al.}~\cite{Dutta_PRA_2007}. In Fig.~\ref{Fig_Comparison}(a) we present the data corresponding to Fig.~2 of \cite{Dutta_PRA_2007}. Our results disagree with theirs in the following significant ways \footnote{We note qualitative similarities between our Fig.~\ref{Fig_Structure3}(b) and Fig.~2 of \cite{Dutta_PRA_2007}, even though our value of $\lambda_z$ is different from that which they quote.}:
{\bf 1.} The results in Fig.~2 \cite{Dutta_PRA_2007} predict stability to  higher dipole strengths than our results  at given trap aspect ratio.
{\bf 2.} We do not find  density-oscillating states for the same  parameters they report. For the density-oscillating region in Fig.~\ref{Fig_Comparison}(a), we only find two-peaked states of the kind shown in Fig.~\ref{Fig_Comparison}(b), and not four-peaked states found in \cite{Dutta_PRA_2007}. We also note that their density-oscillating region extends over a broader range of $\lambda_x$ values than ours.
{\bf 3.} We always find our density-oscillating states to be even with respect to $x$-, $y$-, and $z$-reflections, whereas in Fig.~4 of \cite{Dutta_PRA_2007} ground states are reported which break this symmetry.

The origin of these differences is not clear to us. The imaginary time algorithm used in \cite{Dutta_PRA_2007} to locate ground states is only briefly discussed, although they reported that the results were dependent on the initial (random) states used. Our algorithm (as outlined in Appendix \ref{App_NK}) is built around robust optimization techniques and has been carefully checked against other approaches (e.g.~the cylindrically symmetric results reported in \cite{Ronen_PRA_2006}). We also employ a spherical cutoff interaction potential, which has been shown to improve accuracy on finite grid calculations by minimizing aliasing effects of the long-ranged interaction \cite{Ronen_PRA_2006}.

Another important feature of our work is that we confirm the stability of our solutions by performing a BdG analysis of the excitations. For a stable ground state all the quasi-particles have real energies. However, it is possible to find stationary solutions of the GPE for which some excitations have an imaginary frequency. In this case these excitations would grow exponentially in time and the ground state is dynamically unstable. The importance of this is revealed in Fig.~\ref{Fig_Comparison}(c), where we show a 4-peaked solution of the GPE equation that we obtained at an interaction strength well-above the stability boundary. This solution has excitations with imaginary eigenvalues, so is dynamically unstable. In Fig.~\ref{Fig_Comparison}(d) we compare the energy per particle [Eq.~(\ref{Efn})] of the 4-peaked solution against that of the dynamically stable 2-peaked solution (which ends at the stability boundary, $D\approx10$). Interestingly, the 4-peaked solution exists for dipole strengths where the 2-peaked solution exists, but is of much higher energy. These results demonstrate that BdG solutions are vital in determining stable condensate states.

\subsection{Bogoliubov excitation spectrum: roton modes}\label{Sec_BdGresults}
 \begin{figure}[tbph]
\centering
\includegraphics[width=\columnwidth]{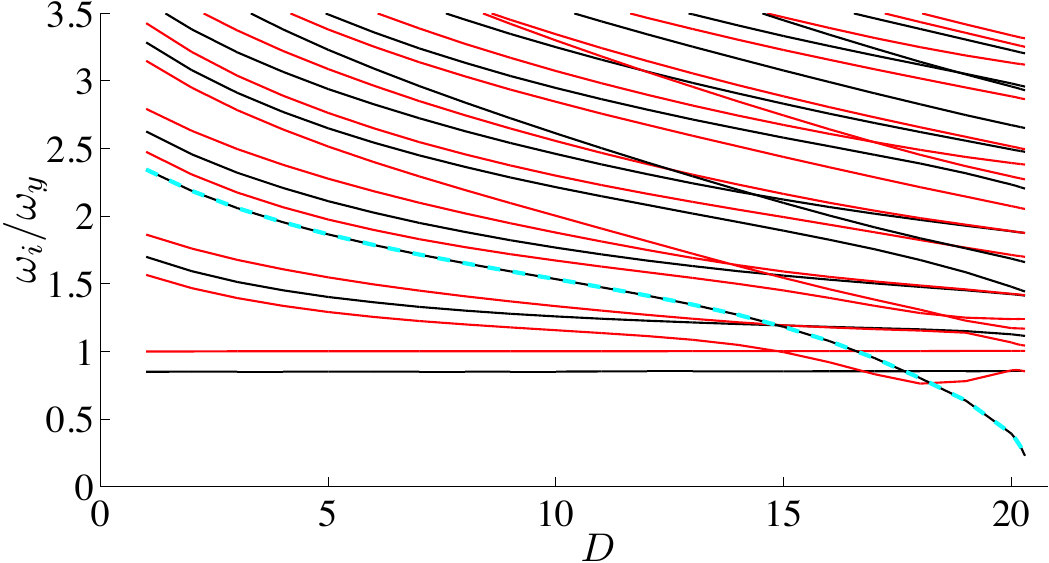}
\caption{(Color online) 
Bogoliubov excitation spectrum as a function of dipole strength $D$, for $\lambda_x=0.85$, $\lambda_z=6$. The softening mode is labelled by the blue dashed line. The colors represent the parity along the $x$-direciton (light/red=even), (dark/black=odd) . For these parameters the condensate is a type-I state up until near collapse where it changes to a type-II state [e.g.~see the relevant phase diagram Fig.~\ref{Fig_Structure3}(b)] } \label{Fig_Bogspec}
\end{figure}

In this section we examine the properties of the quasi-particle excitations of a dipolar condensate. Our particular interest is in the modes that soften and cause the condensate to become dynamically unstable as the dipole strength increases. For example, in Fig.~\ref{Fig_Bogspec} we show the excitation spectrum of a dipolar condensate as a function of the dipole strength. We observe that a mode, which has a relatively high excitation frequency at low values of $D$, decreases its frequency as $D$ increases and eventually approaches zero at the point of instability. Because  this mode tends to have short wavelength features  it is referred to as a rotonic excitation. For definiteness in this section we will refer to the lowest energy such mode for $D$ values close to instability as the roton mode and will label it as the $j=1$ quasi-particle [i.e.~mode  $\{u_1(\mathbf{r}),v_1(\mathbf{r})\}$].  

We will study the properties of these rotons in the same general parameter regime that we used to study the ground states in Sec.~\ref{Sec_GSresults}, i.e.~with $\omega_z\gg \omega_x,\omega_y$. 
 In this regime the roton mode is structureless in the $z$-direction  and exhibits structure in the $xy$-plane. Indeed, the analysis of a uniform quasi-two-dimensional dipolar condensate confined tightly along the $z$ direction predicts that the roton modes lie in the $xy$-plane with a characteristic momentum set by the $z$-confinement length scale \cite{Fischer_PRA_2006}.
The study of roton modes in a cylindrically symmetric pancake trap ($\omega_z\gg \omega_x=\omega_y$) showed that its structure revealed properties of the condensate state \cite{Ronen_PRL_2007}. Noting that for the cylindrically-symmetric trap quasi-particles are eigenstates of  the angular momentum operator $\hat{L}_z$ with eigenvalue $m_z$, two cases of roton modes were found: when the condensate was in a normal (type I) ground state  a \textit{radial roton} emerged with  $m_z=0$. When the condensate was in a biconcave  (type V) state an \textit{angular roton} with $|m_z|>0$ emerged.

Our primary concern here is to examine the structure of the roton modes in the fully anisotropic trap, and the relationship this has to the various types of condensate ground state. 
We begin by introducing the techniques we use to visualize and characterize the roton excitations.

 \begin{figure}[tbph!]
\centering
\includegraphics[width=\columnwidth]{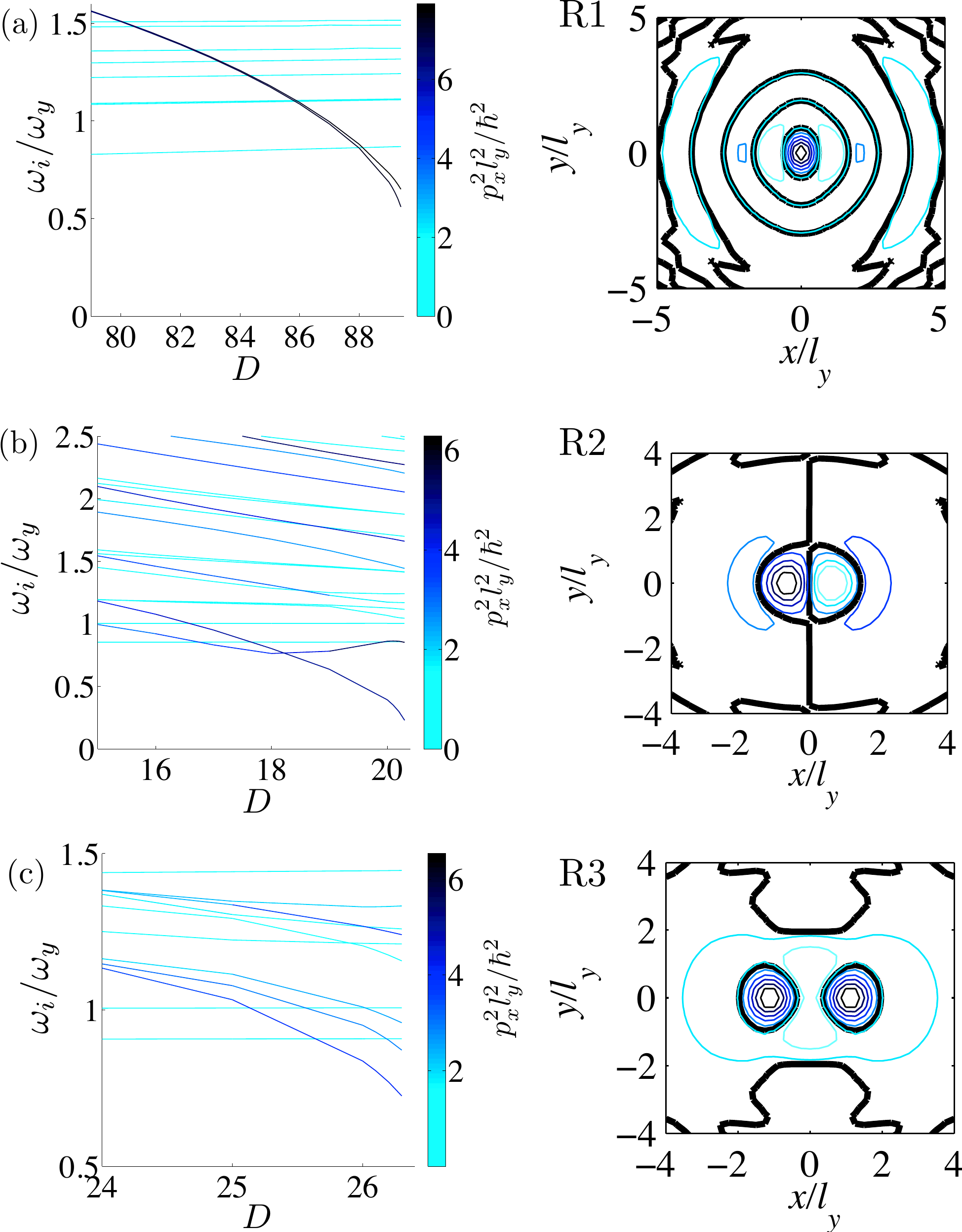}
\caption{(Color online) 
 Bogoliubov excitation spectrum as a function of dipole strength $D$ for (a) $\lambda_x=0.65$, $\lambda_z=9$; (b) $\lambda_x=0.85$, $\lambda_z=6$; (c) $\lambda_x=0.9$, $\lambda_z=6.5$;  Spectra color-coded to indicate $\langle \hat{p}_x^2\rangle$.
 Contours of roton density fluctuations ($\delta n_1$) in the $xy$-plane for (R1)  $D=20.3$ [with type-I condensate]; (R2) $D=26.3$ [type-II]; (R3) $D=89.4$ [type-II]; and trap parameters corresponding to (a), (b), and (c), respectively. Thick lines indicate the nodal lines.
 } \label{Fig_Bog_a}
\end{figure}

\begin{figure}[tbph!]
\centering
\includegraphics[width=\columnwidth]{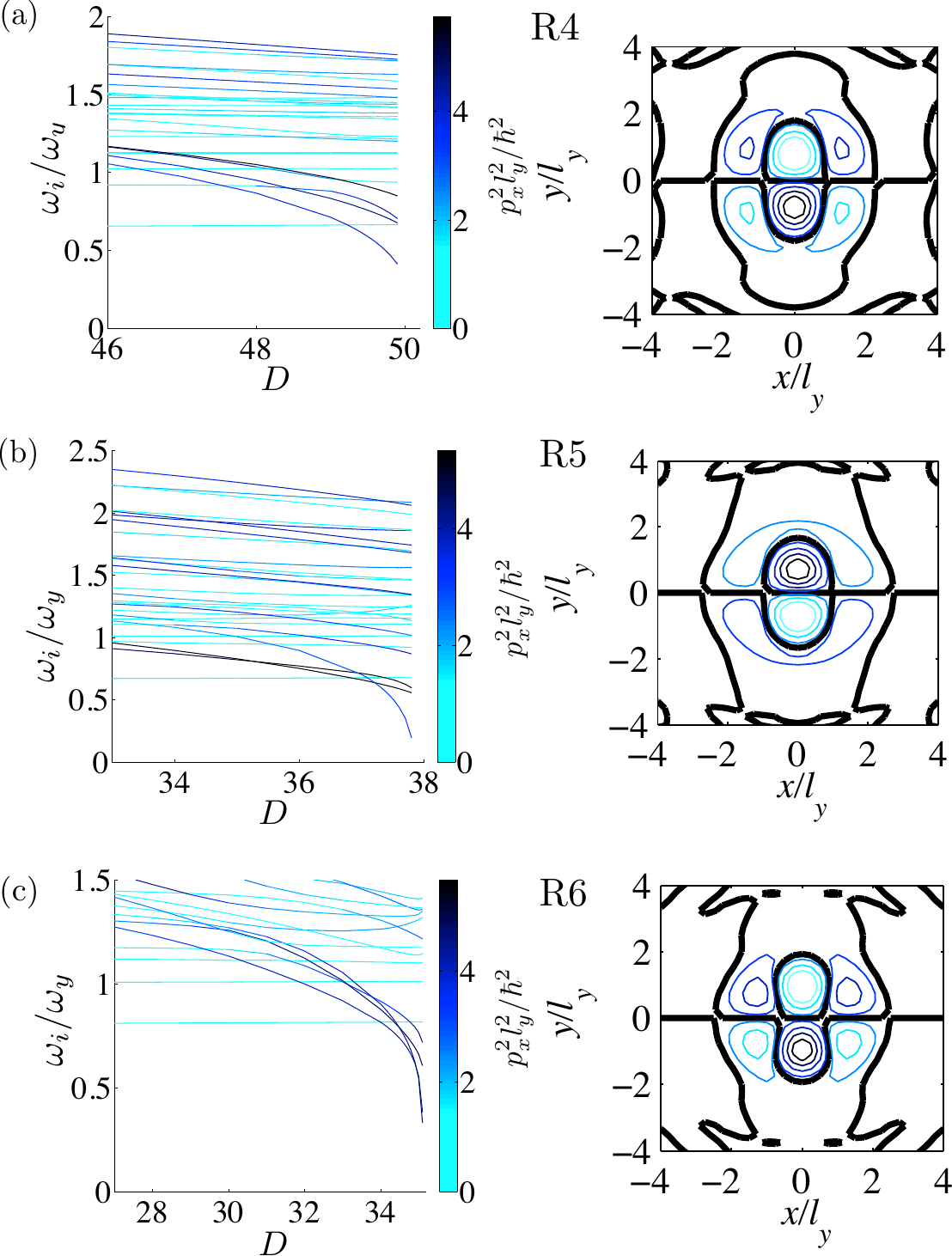}
\caption{(Color online) 
 Bogoliubov excitation spectrum as a function of dipole strength $D$ for (a) $\lambda_x=0.6$, $\lambda_z=6.5$; (b) $\lambda_x=0.65$, $\lambda_z=6$; (c) $\lambda_x=0.8$, $\lambda_z=6.5$;  Contours of roton density fluctuations ($\delta n_1$) in the $xy$-plane for (R4)  $D=50.1$ [with type-III condensate]; (R5) $D=37.8$ [type-IV]; (R6) $D=35.1$ [type-IV]; and trap parameters corresponding to (a), (b), and (c), respectively.  (also see Fig.\ \ref{Fig_Bog_a}). 
 } \label{Fig_Bog_b}
\end{figure}
\subsubsection{Density fluctuation} 
It is useful to consider the effect that the roton, when excited to have a small coherent amplitude $\lambda=|\lambda|e^{i\phi}$ with $|\lambda|\ll1$, has on the condensate. In this case the dynamics of the  total density [i.e.~setting $c_1\to\lambda$ and all other $c_j$ set to zero in Eq.~(\ref{linexpn})]   is given by  \cite{Movies}
\begin{equation}
n(\mathbf{r},t)=N|\psi(\mathbf{r})|^2+2|\lambda|\cos(\omega_1 t-\phi)\sqrt{N}\psi(\mathbf{r})\,\delta n_1(\mathbf{r}),
\end{equation}
  to linear order in $u_1$ and $v_1$.
We have introduced
\begin{equation}
\delta n_j(\mathbf{r})=u_j(\mathbf{r})+v_j(\mathbf{r}),
\end{equation}
as the density fluctuation associated with the $j$-th quasi-particle.
To visualize the roton mode we plot contours of $\delta n_1(\mathbf{x})$ (see Figs.~\ref{Fig_Bog_a} and  \ref{Fig_Bog_b}).

\subsubsection{Roton  characterization}
We also consider how to generalize the qualitative description of the roton modes (e.g.~radial and angular rotons of \cite{Ronen_PRL_2007}) to the anisotropic trap. Because the excitations are not (in general) eigenstates of $\hat{L}_z$ this characterization cannot be performed by inspection of the $m_z$ value of the relevant mode. 
Thus, we propose to  characterize quantitatively the rotonic modes by computing the expectations of various operators, as follows:
\begin{equation}
\langle \hat{O} \rangle_j = \frac{\int d\mathbf{r}\,u^{*}_j(\mathbf{r})\hat{O}u_j(\mathbf{r}) +\int d\mathbf{r}\,v^{*}_j(\mathbf{r})\hat{O}v_j(\mathbf{r}) }{\int d\mathbf{r}\,\left[|u_j(\mathbf{r})|^2+|v_j(\mathbf{r})|^2\right]},
\end{equation}
with $\hat{O}$ being $\hat{p}_x^2$,  $\hat{p}_y^2$, or $\hat{L}_z^2$ (where $\hat{p}_x=-i\hbar\frac{\partial}{\partial x}$ etc.). 
The  results of this analysis for the roton modes are given in Table~\ref{TabRotonData}, and the $\langle \hat{p}_x^2\rangle$ values for all quasi-particles are used to color-code spectra in Figs.~\ref{Fig_Bog_a} and \ref{Fig_Bog_b}. We note that Wilson \textit{et al.}~\cite{Wilson_PRL_2010} used a similar procedure to assign  a momentum to each quasiparticle according to  $p_j=\sqrt{\langle \hat{p}_{x}^2\rangle_j}$, and allow them to approximately extract a dispersion relation in the cylindrically trapped gas. We interpret  these expectations as follows: 
\begin{itemize}
\item{\textit{Angular character}:} We take $\langle\hat{L}_z^2\rangle\gtrsim4\hbar^2$ to define the angular characteristic of a roton mode. Note: We take an $\langle\hat{L}_z^2\rangle$ value consistent with $|m_z|=2$  to define the angular characteristic because this is the minimum value of angular momentum found for an angular roton in the cylindrically symmetric case.
\item{\textit{Linear character}:} When $\langle\hat{p}_x^2\rangle> \langle\hat{p}_y^2\rangle$ the roton mode oscillates more rapidly along the $x$ direction and we refer to the mode as a linear roton along $x$. Similarly, when  $\langle\hat{p}_y^2\rangle> \langle\hat{p}_x^2\rangle$ we have a linear roton  along $y$. A radial roton has no directional preference, i.e.~has $\langle\hat{p}_y^2\rangle\approx \langle\hat{p}_x^2\rangle$.
 \end{itemize}
Note that these characteristics are not exclusive, 
for example, the analysis R6 in Table~\ref{TabRotonData} reveals a roton that has both linear and angular characteristics.  
  We emphasize that our characteristics agree with the terminology adopted by Ronen \textit{et al.}~\cite{Ronen_PRL_2007} for the cylindrically-symmetric case (e.g.~cases R7 and R8 in Table~\ref{TabRotonData}): R7 has angular character (termed an angular roton in \cite{Ronen_PRL_2007}) for a (type-V) biconcave condensate. 
  R8 is a radial roton for a cylindrically symmetric  normal  (type-I) condensate. We note that in the cylindrically symmetric case  we must have $\langle\hat{p}_x^2\rangle=\langle\hat{p}_y^2\rangle$, thus there can be no linear character.
  
\subsubsection{Roton analysis}
 In Fig.~\ref{Fig_Bog_a} we give some examples of roton modes for normal (type-I) condensates  and density-oscillating condensates with peaks along the $x$-direction (type-II). The quantitative analysis and characterization of these is given as results R1-R3 in Table~\ref{TabRotonData}. For these examples the roton varies rapidly along the $x$-direction. More generally, the spectra shown in Fig.~\ref{Fig_Bog_a} reveal that, of the low-energy quasi-particle modes considered, the modes that most rapidly descend as $D$ increases are those with the largest values of $\langle\hat{p}_x^2\rangle$. We also note that the roton for the two type-II states considered differ in their effect (as a density perturbation) on the condensate: in one case [Fig.~\ref{Fig_Bog_a} R2]  the roton causes the two peaks of the condensate to oscillate out-of-phase, while the other case [Fig.~\ref{Fig_Bog_a} R3] causes these peaks to oscillate in-phase  (also see \cite{Movies}). In contrast, for the normal ground state the density fluctuation of the roton is strongest at trap center [Fig.~\ref{Fig_Bog_a} R1] where the ground state has its peak density.

 In Fig.~\ref{Fig_Bog_b} we consider cases in which the condensate has structure in the $y$-direction, i.e.~type-III and type-IV condensates. The quantitative analysis and characterization of these is given as results R4-R6 in Table~\ref{TabRotonData}.
 Interestingly, the emergence of $y$ peaks does not mean that the roton will be most rapidly varying along the $y$-direction, e.g.~as revealed in the momentum expectations of R4 in Table~\ref{TabRotonData}. However, we do observe that the additional $y$ structure tends to emerge in the roton (see $\delta n_1$ plots in Fig.~\ref{Fig_Bog_b}), and is accompanied by an increase in $\langle \hat{p}_y^2\rangle$. The spectra for these states in Fig.~\ref{Fig_Bog_b} indicate that modes with the highest values of $\langle \hat{p}_x^2\rangle$ do not preferentially go soft first.
 The density fluctuation maxima associated with these rotons generally occur at locations corresponding to the peak density of the condensate, and dynamically causes the peaks oscillate in various ways (e.g.~see \cite{Movies}).
 
 A general trend we see is across all states considered is that when the condensate has some biconcavity in its structure (as discussed Sec.~\ref{sec:gstypes}), that the roton tends to exhibit angular character (see Table~\ref{TabRotonData}).

\begin{widetext}
\begin{table*}[bpht!]
\begin{tabular*}{0.95\textwidth}{@{\extracolsep{\fill}}  p{0.85cm}  p{2.05cm}  | c   c c   |    p{1.35cm}   p{1.35cm}   p{1.35cm}   c  c  c}
    \hline \hline
\multicolumn{2}{c|}{ } & \multicolumn{3}{ c| }{Condensate} & \multicolumn{5}{ c}{Roton} \\ \hline
Case & 
\begin{minipage}{1.5cm}\vspace*{-0.1cm} Parameters $(\lambda_x,\lambda_z,D)$ \end{minipage}
& Figure & Type & biconcave 
& \begin{minipage}{1.0cm}\vspace*{-0.3cm}  \[\frac{\langle p_{x}^2\rangle_1 l_y^2}{\hbar^2}\]\end{minipage}
&  \begin{minipage}{1.0cm}\vspace*{-0.3cm}  \[\frac{\langle p_{y}^2\rangle_1 l_y^2}{\hbar^2}\]  \end{minipage}
&  \begin{minipage}{0.85cm}\vspace*{-0.3cm}  \[\frac{\langle L_z^2\rangle_1}{ \hbar^2  }\] \end{minipage} 
& \begin{minipage}{1.5cm}\vspace*{-0.1cm} Linear character\end{minipage} 
& \begin{minipage}{1.5cm}\vspace*{-0.1cm} Angular character\end{minipage}    &\\ \hline 
 \multicolumn{10}{ c  }{rotons in a fully anisotropic trap} &\\ \hline
R1 & $(0.65, 9, 89.4)$ & \ref{Fig_StructureZoom}I & I & \tickNo & $ 7.4$  &$ 2.8$& $0.6$  & \tickYes & \tickNo & \\
R2 & $(0.85,6 ,20.3)$ & \ref{Fig_StructureZoom}II(a), \ref{Fig_Structure2}(f) & II & \tickNo &  $ 4.9$  &$ 2.5$ &$1.2$ & \tickYes & \tickNo & \\
R3 & $(0.9,6.5,26.3)$ & \ref{Fig_StructureZoom}II(b) & II & \tickYes & $ 4.2$  &$ 3.1$&$ 4.2$ & \tickYes & \tickYes & \\
R4 & $(0.6,6.5,50.1)$ & \ref{Fig_StructureZoom}III, \ref{Fig_Structure2}(d) & III &  {\tiny{\tickYes}} & $4.3$ & $ 3.2$&$ 3.7$ & \tickYes & {\tiny{\tickYes}} & \\
R5 & $(0.65,6,37.8)$  & \ref{Fig_StructureZoom}IV, \ref{Fig_Structure2}(e) & IV & \tickNo & $ 3.3$  & $ 3.9$ &$ 2.3$ & \tickYes & \tickNo &\\
R6 & $(0.8,6.5,35.1)$  & \ref{Fig_Structure2}(b)  & IV & \tickYes & $4.8 $  & $ 3.0$&$ 5.6$ & \tickYes & \tickYes &\\ \hline
 \multicolumn{10}{ c  }{rotons in a cylindrically symmetric trap} &\\ \hline
R7 &  $(1,6.5,24)$ &\ref{Fig_StructureZoom}V, \ref{Fig_Structure2}(c) & V & \tickYes & $ 4.0$&$4.0$&4 & \tickNo & \tickYes &    \\
R8 &  $(1,9,48.5)$ & 2(Ia) of \cite{Ronen_PRL_2007} & I & \tickNo & $ 5.0$&$5.0$&0 & \tickNo &  \tickNo  &    \\
    \hline \hline
  \end{tabular*}
     \caption{Roton mode properties for various types of ground states. The small ticks for bi-concavity and angular character qualities of R4 indicate that these are marginal. Excitations for the two cylindrically symmetric cases R7 and R8 are not shown in figures here, however R8 corresponds to Fig.~2(Ic) of \cite{Ronen_PRL_2007}.
     Note that there are two degenerate roton modes for  R7  with $|m_z|>0$, and both have identical squared linear momentum and angular momentum expectations.\label{TabRotonData} }
\end{table*}
\end{widetext}

\section{Conclusions}\label{Sec_Conclusions}
In conclusion, we have mapped the stability and structure of anisotropically trapped dipolar BECs in parameter space, and explored the occurrence of  density-oscillating condensates. The parameter regime we have examined is dominated by ground states with two peaks along the $x$ or $y$ direction. We observe these ground states to form clear sub-regions of parameter space, which branch off in the highly anisotropic case ($\lambda_x\ll1$).

Collapse instability is associated with the softening of a roton excitation, which in anisotropic traps we typically find to have a linear character with it being most highly excited in the direction of weakest trap frequency. The softening roton mode has characteristics associated with the ground state structure, e.g., if the ground state has biconcave character then the roton tends to have an angular character; we also find that the density fluctuations of the roton mode coincide with the peaks of the condensate density.  By increasing the dipole strength slowly until the system is dynamically unstable we would expect that this roton structure will be revealed in the collapse dynamics (e.g.~see \cite{Wilson_PRA_2009}).

The linear character of the roton mode suggests that the critical velocity for  breakdown of superfluidity will be anisotropic in the $xy$-plane. This differs from the anisotropic superfluidity predicted for pancake trapped dipolar condensates in \cite{Ticknor_PRL_2011}, in which the anisotropic arises from tilting the dipole polarization axis into the $xy$-plane. It will be interesting to explore the nature of the rotons in a fully anisotropic trap under much tighter $z$-confinement.
  
As most experiments in dipolar condensates \cite{Lu_PRL_2011,Aikawa_PRL_2012,Lahaye_PRL_2008,Koch_NaturePhys_2008,Muller_PRA_2011,Bismut_unpublished_2012} are conducted in anisotropic traps, we believe our more general study of condensate and roton structure will assist in experiments planning to measure roton properties. 

\section*{Acknowledgments}  This work was supported by the Marsden Fund of New Zealand  contract UOO0924.

\appendix
\section{Brief outline of numerical methods}
In this appendix we briefly discuss the approach we use for solving the GPE and BdG equations. Our approach is similar to that outline in Ref.~\cite{Ronen_PRA_2006}.
We will work in the dimensionless units that were introduced in Sec.~\ref{resultssec}. In these units the single-particle operator and DDI potential take the form
\begin{align}
H_0&=-\frac{1}{2}\nabla^2+\frac{1}{2}\left(\lambda_x^2x^2+y^2+\lambda_z^2z^2\right),\\
V_{ {D}}(\mathrm{r})&={D} \frac{1-3\cos^2\theta}{r^3},
\end{align}
respectively, 
where we have set $NV_{\mathrm{int}}\to V_{D}$ to  explicitly include the condensate number, $N$.

\subsection{Newton Krylov method of obtaining stationary states }\label{App_NK}
GPE solutions can be found by minimizing the energy functional:
\begin{equation} 
E\left[\psi_u \right] =   \frac{1}{\mathcal{N}}\int d\mathbf{r}\, \psi^{*}_u \left[H_0+\frac{1}{2\mathcal{N}}\Phi_{D}(\mathbf{r})   \right]\psi_u,\label{Efn}
\end{equation} 
where
\begin{equation}
\Phi_D(\mathbf{r})=\int d\mathbf{r^{\prime}}\,V_{D}(\mathbf{r}-\mathbf{r}')n_u(\mathbf{r^{\prime}}),\label{Eq_Phi}
\end{equation}
and $n_u(\mathbf{r})=|\psi_u(\mathbf{r})|^2$. To avoid having to deal with the the condensate's normalization constraint we have allowed the condensate orbital to be unnormalized (denoted $\psi_u$) and explicitly scaled out the normalization dependence in $E\left[\psi_u \right]$ using the normalization functional $\mathcal{N}= \int |\psi_u(\mathbf{r})|^2 d\mathbf{r}$ \cite{Modugno_EPJD_2003}. This means that the un-normalized ($\psi_u$) and normalized ($\psi=\psi_u/\sqrt{\mathcal{N}}$) orbitals give the same energy in Eq.~(\ref{Efn}).
We also note that  $\Phi_D(\mathbf{r})$ can be efficiently evaluated using the convolution theorem $\Phi_D(\mathbf{r})=\mathcal{F}^{-1}[\tilde{V}_D(\mathbf{k})\tilde{n}_u(\mathbf{k})],$
where $\mathcal{F}$ is a three-dimensional Fourier transform,  $\tilde{n}_u(\mathbf{k})=\mathcal{F}\{n_u(\mathbf{r})\}$ and $\tilde V_D(\mathbf{r})=\mathcal{F}\{V_D(\mathbf{r})\}$ (which can be evaluated analytically \cite{Goral_PRA_2002,Ronen_PRA_2006}). In our code $\mathcal{F}$ is implemented using forward and reverse fast Fourier transform (FFT) algorithms. The FFT implicitly treats the system as a 3D lattice of condensates of period $2R$, where our numerical grid is cubic with extent $[-R, R]$ in the $x,y,z$ directions and our results here use $R=8$ with 64 points in each direction. We adopt a radial truncation of $V_D(\mathbf{r})$, such that $V_D(\mathbf{r})=0$ for $r>R$, which prevents its overlap with the condensate `copies', giving improved numerical accuracy of $\Phi_D(\mathbf{r})$. The Fourier transform of this truncated potential has the analytic form \cite{Ronen_PRA_2006}
\begin{equation}
\tilde{V}_D^{\mathrm{tr}}(\mathbf{k})=\frac{4\pi D}{3}\left[1+3\frac{\cos(Rk)}{R^2k^2}-3\frac{\sin(Rk)}{R^3 k^3} \right]\left( 3 \cos^2\alpha -1\right),
\end{equation}
where $\alpha$ is the angle between $\mathbf{k}$ and the $k_z$-axis.

 Our general procedure discussed here is similar to that presented in \cite{Ronen_PRA_2006}. However, we differ from that most significantly in that instead of using a conjugate gradient technique we  use a Newton-Krylov algorithm \cite{Kelley_Book_2003} to solve for the condensate. This algorithm iteratively finds zeros of the residual: $\partial E/\partial \psi^*_u=\left[\mathcal{L}_{\mathrm{GP}}\psi_u-\mu\psi_u \right]/\sqrt{\mathcal{N}}$, where
\begin{equation}
\mathcal{L}_{\mathrm{GP}} =H_0 + {\mathcal{N}}^{-1}\Phi_D(\mathbf{r}),\label{LGPnum}
\end{equation} and $\mu=\int d\mathbf{r}\,\psi_u^*\mathcal{L}_{\mathrm{GP}}\psi_u/\mathcal{N}$.
 
 \subsection{Solution of Bogoliubov-de Gennes equations }\label{App_BdG}
We formulate the BdG matrix $\mathcal{L}$ [Eq.\ (\ref{Eq_LBdG})] in the GPE basis, i.e., eigenstates of Eq.\ (\ref{LGPnum}), which we calculate directly on our numerical grid using the Arnoldi algorithm provided by the MATLAB routine `eigs'. The integrals involving the dipolar interaction potential are done analogously to those described in Sec. \ref{Sec_condensate}.
We diagonalize the resulting BdG matrix using the MATLAB routine `eig'. For most cases studied in this paper, we find 400 basis vectors sufficient for convergence of the excitation energies $\omega_i$, however, bases of more than 1200 vectors are required for condensates tightly trapped in the $z$ direction (e.g., for $\lambda_z=9$).

\end{document}